\documentclass[prd,twocolumn,showpacs,preprintnumbers,amsmath,amssymb,superscriptaddress,floatfix,nofootinbib]{revtex4}

\usepackage{graphicx}
\usepackage{epstopdf}
\usepackage{bm}
\usepackage{amsmath}
\usepackage{amsfonts}
\usepackage{amssymb}
\usepackage{color}
\usepackage{multirow}
\usepackage{footnote}
\usepackage{siunitx}
\usepackage{physics}

\usepackage[colorlinks, citecolor=blue,anchorcolor=red,menucolor=red,linkcolor=red,filecolor=red,runcolor=red,urlcolor=blue,frenchlinks=red]{hyperref}
\begin{document}
\title{Bottomonium meson spectrum with quenched and unquenched quark models}

\author{M. Atif Sultan}%
\email{atifsultan.chep@pu.edu.pk}
\affiliation{School of Physics, Nankai University, Tianjin 300071, China}
\affiliation{Centre  For  High  Energy  Physics,  University  of  the  Punjab,  Lahore  (54590),  Pakistan}

\author{Wei Hao}
\email{haowei@nankai.edu.cn}
\affiliation{School of Physics, Nankai University, Tianjin 300071, China}

\author{E. S. Swanson}%
\email{swansone@pitt.edu}
\affiliation{Department of Physics and Astronomy, University of Pittsburgh, Pittsburgh PA 15260, USA.}

\author{Lei Chang}%
\email{leichang@nankai.edu.cn}
\affiliation{School of Physics, Nankai University, Tianjin 300071, China}


\begin{abstract}
An open question in hadronic phenomenology concerns the ``unquenching" effects of higher Fock space components on the leading Fock space description of hadrons. We address this by making a
comparison of the bottomonium spectrum as computed with the relativized Godfrey-Isgur quark model and an unquenched coupled channel model driven by the ``$^3P_0$" mechanism of hadronic decay.
Our results show that both models can describe the spectrum well, indicating that the influence of coupled channel effects can be largely absorbed into the parameters of the quenched quark model. This conclusion is reinforced by a perturbative calculation that shows that the spin-dependence of mass splittings due to mixing with the continuum recapitulates quenched quark model spin-dependent interactions. We also show that softening of the quark-antiquark wavefunction due to continuum mixing improves the description of vector bottomonium decay constants. Together, these results illustrate and
substantiate the surprising robustness of simple constituent quark model descriptions of hadrons.
\end{abstract}

\maketitle

\section{Introduction}

Experimental effort over the past decade has filled out the low-lying bottomonium spectrum~\cite{PDG:2023}. Known states  include the $\Upsilon$ family $\Upsilon(1S,2S,3S,4S)$, $\Upsilon(10753)$, $\Upsilon(10860)$, $\Upsilon(11020)$, $\Upsilon_2(1D)$, the $\eta_b$ family, $\eta_b(1S,2S)$, the $\chi_{bJ}$ family $\chi_{b0}(1P,2P)$, $\chi_{b1}(1P,2P,3P)$, $\chi_{b2}(1P,2P,3P)$, and the $h_b$ family $h_b(1P,2P)$; reviews may be found in Refs.~\cite{Segovia:2016xqb,Rosner:2006jz,Patrignani:2012an}. Here we seek to use this wealth of information to address the effect of the coupling of bare quark-antiquark states to the continuum.

Valence quark models have been widely used in the study of hadron spectroscopy for many years~\cite{DeRujula:1975qlm,Eichten:1978tg,Stanley:1980zm,Godfrey:1985xj,Vijande:2004he,Lakhina:2006fy}.  Generally, these models are regarded as ``quenched'', meaning that the hadrons are described by their lowest Fock components: quark and antiquark ($q\bar{q}$) for mesons and three quarks ($qqq$) for baryons. The models can otherwise differ in the choice of
interactions, computational techniques, and relativistic corrections. Among these models, Godfrey and Isgur’s relativized quark model~\cite{Godfrey:1985xj} (GI model) has become something of a standard due to its broad success in describing mesons of all flavors~\cite{Vijande:2004he,Pantaleone:1985uf,Lakhina:2006fy,Ebert:2009ua}.

However, the emergence of a plethora of states that are inconsistent with the predictions of traditional quark models has raised new challenges in hadronic spectroscopy.
To deal with this, authors have sought to expand the constituent quark model in various ways. One way forward is the incorporation of gluoninc degrees of freedom~\cite{Dudek:2008sz,USQCD:2022mmc,Shepherd:2016dni}, which must become evident in the spectrum approximately 1 GeV above flavor threshold. Loop effects in the interquark interaction have also been considered and shown to be relevant in certain open flavor systems~\cite{Lakhina:2006fy}. Novel diquark contributions to  spectroscopy have also been considered~\cite{Giron:2021sla,Giron:2020qpb,Giron:2020fvd,Giron:2019bcs,Giron:2019cfc}. Lastly, the prevalence of states near thresholds have led to the consideration of screening effects in the potential~\cite{Song:2015nia,Song:2015fha,Wang:2018rjg,Wang:2019mhs,Hao:2019fjg,Hao:2022ibj,Feng:2022esz,Kanwal:2022ani,Swanson:2005rc}.  A similar approach considers the effect of direct coupling to the hadronic continuum, which we call ``unquenching"~\cite{Ferretti:2012zz,Ferretti:2013faa,Ferretti:2013vua,Yang:2023tvc,Hao:2022vwt,vanBeveren:2020eis,Rupp:2017iat,Rupp:2016uee,Thomas:2014dpa}.
Indeed, one might expect that screening and unquenching are related since they approximately embody the same effects of long range vacuum polarization~\cite{Li:2009ad}. At minimum, quark pair creation effects give rise to
virtual $q\bar{q} - q\bar{q}$ ($qqq - q\bar{q}$) components in the hadron wave function and shift  the physical mass with respect to the bare mass~\cite{Patrignani:2012an,Santopinto:2014fya}. Such mixing can  strongly influenced by nearby channels~\cite{Patrignani:2012an, Isgur:1999cd}. An extensive literature exploring these effects in mesons~\cite{Kalashnikova:2005ui,Danilkin:2009hr,Ferretti:2012zz,Ortega:2009hj,Lu:2016mbb}, and  baryons~\cite{Brack:1987dg,Silvestre-Brac:1991qqx,Fujiwara:1992yv} exists.

Incorporating coupled channel effects in a constituent quark model requires an explicit realization of quark pair creation. In this work we employ the ``$^3P_0$ model", which has
been widely used to calculate the strong decays~\cite{Song:2015nia,Song:2015fha,Lu:2016bbk,Wang:2018rjg,Wang:2019mhs,Hao:2019fjg} and coupled-channel effects~\cite{Kalashnikova:2005ui,Danilkin:2009hr,Ferretti:2012zz,Ortega:2009hj,Lu:2016mbb}. It is common to approximate the mesonic wavefunctions in decay calculations as simple harmonic oscillator (SHO) wavefunctions due to the complexity of the formalism~\cite{Ferretti:2012zz,Ferretti:2013faa,Santopinto:2014fya,Wang:2017pxm}. While generally successful, this approximation can introduce additional uncertainty, especially when different flavor quarks are present.
To avoid this we use numerical wave functions obtained from diagonalizing the model in the $q\bar q$ sector. This approach has also been adopted in other work, such as
Refs.~\cite{Lu:2016mbb,Lu:2016bbk,Hao:2019fjg}.

The main purpose of this work is to compare the efficacy of a  quenched and an unquenched constituent quark model in describing the bottomonium spectrum. As a corollary, we seek to determine if quark model assignments for experimental states can differ in the two approaches. We are thus fortunate that the rich experimental information available permits determining model parameters in both cases with reasonable accuracy.

We choose to use the Godfrey-Isgur relativized quark model for the unquenched work. As discussed, this model has enjoyed wide success and use since it provides an accurate meson spectrum over all quark flavors. The model is also used to provide the bare $q\bar q$ states for the unquenched model, with channel coupling driven by the $^3P_0$ $q\bar q$ creation vertex, as mentioned above. Computations are typically done by saturating the intermediate $qq\bar q\bar q$ with meson pairs. In practice
~\cite{Ferretti:2013faa,Ferretti:2012zz,Ferretti:2013vua,Lu:2016mbb}, only low-lying open bottom mesons are considered, specifically combinations of S-wave $B$ and $B_s$ mesons. In principle, this sum should extend over all pairs of meson quantum numbers, and may not even be well-defined (this remains an open topic of investigation). Here we extend the computation by including excited S-waves and P-waves mesons in the intermediate state. Our work is closest in goals and form to Ref.~\cite{Lu:2016mbb}, which exaimines the effect of unquenching on spectroscopic mixing in bottomonium, and Ref.~\cite{Kanwal:2022ani}, which examines the unquenched charmonium spectrum.

This work is organized as follows. In  Sec.~\ref{sec:model}, we give a brief review of the  relativistic GI model, the coupled channel model, and the $^3P_0$ model. In Sec.~\ref{sec:results}, the numerical results are presented. The summary is given in Sec. \ref{sec:summary}.

\section{The model}
\label{sec:model}

\subsection{The quenched quark model}
\label{sec:GI}

The GI model~\cite{Godfrey:1985xj} is defined by a Hamiltonian of the form
\begin{eqnarray}
H_A&=&(p^2+m_1^2)^{1/2}+(p^2+m_{2}^2)^{1/2} +  \tilde{H}_{12}^{\text{conf}} \nonumber \\
&&+\tilde{H}_{12}^{\text{hyp}}+\tilde{H}_{12}^{\text{so}},
\label{ha}
\end{eqnarray}
where $\tilde{H}^\text{conf}_{12}$ is a spin-independent confinement term, $\tilde{H}^\text{hyp}_{12}$ is the color-hyperfine tensor and contact interactions, and
$\tilde{H}^\text{so}_{12}$ incorporates vector and scalar spin-orbit interactions.

Motivated by the nonrelativistic expansion of quark field operators, relativistic effects are introduced by smearing in coordinate space and by including momentum-dependent factors.
The smearing transformation can be expressed as
\begin{eqnarray}
    \tilde{f}(r) &=& \int f(r)  \rho(\boldsymbol{r}-\boldsymbol{r}')\, dr'^3, \nonumber \\
     \rho(\boldsymbol{r}-\boldsymbol{r}') &=&\frac{\sigma^3}{\pi^{3/2}}e^{-\sigma^2(\boldsymbol{r}-\boldsymbol{r}')^2},
\end{eqnarray}
where $\rho(\boldsymbol{r}-\boldsymbol{r}')$ is the smearing function and $\sigma$ is a parameter to be fit to experiment. In the case of the confiment interaction, the function $f(r) = H^{\text{conf}}$ is taken to be
\begin{eqnarray}
 H_{12}^{\text{conf}}  &=&G(r)+S(r), \nonumber \\
 G(r)&=&\frac{\alpha_s(r)}{r} \boldsymbol{F_1} \cdot \boldsymbol{F_2},\\
 S(r)&=& (br+c) \,  \boldsymbol{F_1} \cdot \boldsymbol{F_2},
 \label{hyp}
\end{eqnarray}
where the $\alpha_s(r)$ is a fixed model of the running coupling constant of QCD as specified in Ref.~\cite{Godfrey:1985xj}, $b$ is the string tension, and $c$ is constant. $\boldsymbol{F}$ is related to the Gell-Mann matrix by $\boldsymbol{F}_1=\boldsymbol{\lambda}_1/2$ and $\boldsymbol{F}_2=-\boldsymbol{\lambda}^*_2/2$ for quark and antiquark, respectively, with $\langle\boldsymbol{F}_1\cdot\boldsymbol{F}_2\rangle=-4/3$ for mesons.

Momentum-dependence is introduced as factors:
\begin{eqnarray}
\tilde{G}(r) &\to& \left(1+\frac{p^2}{E_1 E_2}\right)^{1/2} \tilde{G}(r) \left( 1+\frac{p^2}{ E_1 E_2}\right)^{1/2},\nonumber\\
\frac{\tilde{V}_i(r)}{m_1 m_2} &\to& \left( \frac{m_1 m_2}{E_1 E_2}\right)^{1/2+\epsilon_i}  \frac{\tilde{V}_i(r)}{m_1 m_2} \left( \frac{m_1 m_2}{E_1 E_2}\right)^{1/2+\epsilon_i},\label{eqGV}
\end{eqnarray}
where $E_1=(p^2+m^2_1)^{1/2}$, $E_2=(p^2+m^2_2)^{1/2}$ are the energies of the quark and antiquark in
the meson, respectively. The index \textit{i} that appears in $\tilde{V}_i(r)$ and $\epsilon_i$ runs over the contact(c), tensor(t), vector spin-orbit[so(v)], and scalar spin-orbit[so(s)] potentials.

Finally, the base hyperfine interaction is written as
\begin{eqnarray}
H^{\text{hyp}}_{12} &=& -\frac{\alpha_s(r)}{m_1m_2}\Bigg[\frac{8\pi}{3}\boldsymbol{S}_1\cdot\boldsymbol{S}_2\delta^3 (\boldsymbol r) + \nonumber \\
&& \frac{1}{r^3} \Big(\frac{3\boldsymbol{S}_1\cdot\boldsymbol{r} \boldsymbol{S}_2\cdot\boldsymbol{r}}{r^2}
-\boldsymbol{S}_1\cdot\boldsymbol{S}_2\Big)\Bigg] \boldsymbol{F_1} \cdot \boldsymbol{F_2}.
\end{eqnarray}
The base spin-orbit interaction can be expressed as
\begin{equation}
 H^{\text{so}}_{12}=H^{\text{so(cm)}}_{12}+H^{\text{so(tp)}}_{12},
\end{equation}
where $H^{\text{so(cm)}}_{12}$ is the
color-magnetic term and $H^{\text{so(tp)}}_{12}$ is the Thomas-precession term, i.e.,
\begin{equation}
H^{\text{so(cm)}}_{12} = -\frac{\alpha_s(r)}{r^3}\left(\frac{1}{m_1}+\frac{1}{m_2}\right)\left(\frac{\boldsymbol{S}_1}{m_1}+\frac{\boldsymbol{S}_2}{m_2}\right) \cdot\bm{L}(\boldsymbol{F}_1\cdot\boldsymbol{F}_2),
\end{equation}
\begin{equation}
H^{\text{so(tp)}}_{12}=\frac{-1}{2r}\frac{\partial H^{\text{conf}}}{\partial
r}\Bigg(\frac{\boldsymbol{S}_1}{m^2_1}+\frac{\boldsymbol{S}_2}{m^2_2}\Bigg)\cdot \boldsymbol{L},
\end{equation}
where $\boldsymbol{S}_1$ and $\boldsymbol{S}_2$ represent the spin of the quark and antiquark respectively, and $\boldsymbol{L}$ is the orbital momentum between quark and antiquark.

\subsection{The unquenched quark model}
\label{sec:couple}

Our unquenched model is defined by restricting the Fock space expansion relevant to a given hadron to the two lowest, $q\bar q$ and $qq\bar q \bar q$, components~\cite{Swanson:2005rc,Ferretti:2013faa,Ferretti:2012zz,Ferretti:2013vua,Ortega:2016mms,Li:2009ad,Lu:2016mbb}. The model Hamiltonian can be written as
\begin{equation}
	H = H_A+H_{BC}+H_I,
\end{equation}
where $H_A$ describes the bare state with mass $M_0$, which is calculated in the GI model. $H_{BC}$ is the sum of the kinetic energies of $B$ and $C$, \textit{i.e.} $E_{BC} =\sqrt{m_B^2 +p^2} + \sqrt{m_C^2 +p^2}$ (final state interactions in the continuum are neglected), and  $H_I$ describes the coupling between a bare state and a meson pair, to be discussed in the next section.
The physical mass $M$ is defined as

\begin{equation}
\label{m}
M = M_0 + \Delta M,
\end{equation}
with the mass shift given by the perturbative expression

\begin{equation}
\label{dm}
\Delta M = \sum_{BC\ell J} \int_0^{\infty} p^2 dp \mbox{ } \frac{|\langle (BC)_\ell^J;p | H_I | A \rangle |^2}{M - E_B(p)-E_C(p) -i\epsilon},
\end{equation}
where the principle value is used to compute the integral. Of course, the imaginary part of the integral is related to the decay width, $\Gamma(A \to BC) = - 2\Im \Delta M_{BC}$.
Nonperturbative mass shifts can also be considered, but do not have a great impact on the general behavior of the model~\cite{Barnes:2007xu}.
Lastly, the probability that the physical state is in the two-meson channel $BC$ is

\begin{eqnarray}
\label{deltam2}
P_{BC}^{J\ell} &=&\left(1+\sum_{BC\ell J} \int_0^{\infty} p^2 dp \mbox{ }
\frac{|\langle (BC)_\ell^J;p | H_I | A \rangle |^2}{[M_A -
E_B(p)-E_C(p)]^2}\right)^{-1} \nonumber \\
&& \int_0^{\infty} p^2 dp \mbox{ } \frac{|\langle (BC)_\ell^J;p | H_I
| A \rangle |^2}{[M_A - E_B(p)-E_C(p)]^2}.
\end{eqnarray}

Masses to be to fit to experiment are obtained from Eqs.~\ref{m} and~\ref{dm}. The sum in Eq.~\ref{dm} is assumed to saturate when summing over S-waves, excited S-wave, and P-wave intermediate channels:
$B\bar B$, $B\bar B^*$, $B^*\bar B^*$, $B_s\bar B_s^*$, $B_s^*\bar B_s^*$, $B\bar{B}(1^3P_0)$, $B^*\bar{B}(1^3P_0)$, $B\bar{B}(1^1P_1)$, $B^*\bar{B}(1^1P_1)$, $B\bar{B}(1^3P_1)$, $B^*\bar{B}(1^3P_1)$, $B\bar{B}(1^3P_2)$, $B^*\bar{B}(1^3P_2)$, $B\bar{B}(2^1S_0)$, $B\bar{B}(2^1S_0)$, $B^*\bar{B}(2^3S_1)$ and $B^*\bar{B}(2^3S_1)$. The masses used for these mesons in the following computation are listed in Table~\ref{final mass}.

\begin{table*}[htpb]
\caption{$B$ and $B_s$ Masses~\cite{PDG:2023}.}
\label{final mass}
\begin{center}
\begin{tabular}{cccccccccccc}
\hline
\hline
States      &$B^0 $   &$B^+$  &$B^*$  &$B_s$  &$B_s^*$ &$B(2^1S_0)$  &$B(2^3S_1)$  &$B(1^3P_0)$ &$B(1^1P_1)$ &$B(1^3P_1)$ &$B(1^3P_2)$ \\
Mass       &5279.65           &5279.34    &5324.7 &5368.8         &5415.4   &5905         &5934         &5756        &5777        &5784        &5797   \\
\hline
\hline
\end{tabular}
\end{center}
\end{table*}

\subsection{$^3P_0$ model}
\label{sec:3P0}

In our calculations, we used the $^3P_0$ model to evaluate the coupled-channel effects. The quark-antiquark pair-creation operator can be expressed as follows~\cite{Ferretti:2013faa,Ferretti:2012zz,Ferretti:2013vua}:
\begin{equation}
	\label{eqn:Tdag}
	\begin{array}{rcl}
	H_I &=& -3 \, \gamma_0^{eff} \, \int d \boldsymbol{p}_3 \, d \boldsymbol{p}_4 \,
	\delta(\boldsymbol{p}_3 + \boldsymbol{p}_4) \, C_{34} \, F_{34} \,
	{e}^{-r_q^2 (\boldsymbol{p}_3 - \boldsymbol{p}_4)^2/6 }\,  \\
	& & \left[ \chi_{34} \, \times \, {\cal Y}_{1}(\boldsymbol{p}_3 - \boldsymbol{p}_4) \right]^{(0)}_0 \,
	b_3^{\dagger}(\boldsymbol{p}_3) \, d_4^{\dagger}(\boldsymbol{p}_4) ~,
	\end{array}
\end{equation}
where $C_{34}$, $F_{34}$ and $\chi_{34}$ are the color singlet wave function, flavor singlet wave function and spin triplet wave function of the created $q\bar{q}$ pair, respectively. The quark and antiquark creations operators are denoted $b_3^{\dagger}(\boldsymbol{p}_3)$ and $d_4^{\dagger}(\boldsymbol{p}_4)$ respectively. The pair creation strength is  $\gamma_0^{eff}=\frac{m_n}{m_i}\gamma_0$, where $\gamma_0=0.4$. This value has been widely used to study the decay of the mesons~\cite{Ferretti:2013faa,Barnes:2005pb,Barnes:1996ff,Barnes:2002mu}. Here $m_n$ refers to a light constituent quark mass, $m_u=m_d$, while $m_i$ is the constituent mass of the created quarks. In addition, the constituent quark mass of bottom, up/down and strange quark are taken to be 4.977, 0.22
and 0.419 GeV, respectively~\cite{Godfrey:1985xj}. The parameter $r_q$, which describes the quark pair creation smearing, is in the range 0.25 to 0.35 fm~\cite{Silvestre-Brac:1991qqx,Geiger:1991ab,Geiger:1991qe,Geiger:1996re}. We use the value  $r_q = 0.3$ fm in our calculation.

\section{Results and Discussion}
\label{sec:results}

Eighteen well-established bottomonium states were fit in the quenched and unquenched models using the relative error, $R^2$, as the objective function:

\begin{equation}
 R^2=\frac{1}{N}\sum_i^N \left(\frac{M_\text{expt}(i)-M_\text{thy}(i)}{M_\text{expt}(i)}\right)^2,
 \end{equation}
The experimental masses used in the fit are listed column three of Tables~\ref{tab:mass1} and \ref{tab:mass2}.

The parameters obtained for both fits are reported in Table~\ref{gipar}. We see that the largest shift is in the constant term, as might be expected if unquenching effects are largely quantum number independent (we discuss this further below). The bottom quark mass is reduced by approximately 8\%, indicating that a more diffuse wavefunction is required to offset the effects of coupling to the continuum. We note that the strong coupling model was not modified in the unquenched calculation. This is a strong assumption, which, in the case of general renormalization, is incorrect. However, Geiger and Isgur have argued that the dominate effect of unquenching is to induce a shift in the linear part of the effective interaction~\cite{Geiger:1989yc}. That we obtain a good unquenched spectrum while holding the Coulombic interaction fixed supports this claim.

 Table~\ref{tab:mass1} shows the resulting quenched spectrum, along with experimental masses, and, for comparison, fit results from five other models. Evidently, a good fit is obtained with an average relative error of 0.16\% and an average absolute error of 1.3 MeV.  Table~\ref{tab:mass2} gives the result of the unquenched model, with the bare mass in column 4 and the physical mass in column 6. Once again, the fit is of high quality, with an average relative error of 0.21\% and an average absolute error of 2.1 MeV. The similarity of the quenched and unquenched spectra demonstrates that the majority of the effect of unquenching can be absorbed by the model parameters. This observation underpins the general robustness of the constituent quark model.

\begin{table*}[htpb]
\begin{center}
\caption{\label{tab:mass1} Experimental and Theoretical Quenched Bottomonium Spectra (MeV).}
\footnotesize
\begin{tabular}{cccc|cccccc}
\hline\hline
  $n^{2S+1}L_J$  & States               & RPP~\cite{PDG:2023}      &$M$       &GI~\cite{Godfrey:1985xj}   &YL~\cite{Lu:2016mbb}     &JFL~\cite{Liu:2011yp}   &JF~\cite{Ferretti:2013vua}    &MGI~\cite{Wang:2018rjg}  \\\hline
  $1^1S_0$     & $\eta_b(1S)$           &$9398.7\pm2.0$            &9423.54   &9394   &9395   &9392  &9391  &9398    \\
  $1^3S_1$     & $\Upsilon(1S)$         &$9460.30\pm0.26$          &9486.25   &9459   &9459   &9460  &9489  &9463    \\
  $2^1S_0$     & $\eta_b(2S)$           &$9999\pm4$                &9985.53   &9975   &9982   &10005 &9980  &9989    \\
  $2^3S_1$     & $\Upsilon(2S)$         &$10023.26\pm0.31$         &10013.16  &10004  &10011  &10026 &10022 &10017    \\
  $3^1S_0$     & $\eta_b(3S)$           &                          &10325.80  &10333  &10353  &10338 &10338 &10336     \\
  $3^3S_1$     & $\Upsilon(3S)$         &$10355.2\pm0.5$           &10345.58  &10354  &10373  &10352 &10358 &10356    \\
  $4^1S_0$     & $\eta_b(4S)$           &                          &10593.40  &10616  &       &10593 &      &10597    \\
  $4^3S_1$     & $\Upsilon(4S)$         &$10579.4\pm1.2$           &10609.44  &10633  &10654  &10603 &      &10612   \\
  $5^1S_0$     & $\eta_b(5S)$           &                          &10829.50  &10860  &       &10813 &      &10810   \\
  $5^3S_1$     & $\Upsilon(5S)$         &$10885.2^{+2.6}_{-1.6}$   &10845.30  &10875  &10999  &10820 &      &10822   \\
  $6^1S_0$     & $\eta_b(6S)$           &                          &11006.60  &11079  &       &11008 &      &10991   \\
  $6^3S_1$     & $\Upsilon(6S)$         &$11000\pm4$               &11015.72  &11092  &11265  &11023 &      &11001   \\
  $1^3P_0$     & $\chi_{b0}(1P)$        &$9859.44\pm0.42\pm0.31$   &9863.24   &9845   &9851   &9875  &9849  &9858   \\
  $1^1P_1$     & $h_b(1P)$             &$9899.3\pm0.8$             &9897.06   &9881   &9886   &9916  &9885  &9894   \\
  $1^3P_1$     & $\chi_{b1}(1P)$        &$9892.78\pm0.26\pm0.31$   &9891.50   &9875   &9880   &9907  &9879  &9889   \\
  $1^3P_2$     & $\chi_{b2}(1P)$        &$9912.21\pm0.26\pm0.31$   &9911.36   &9896   &9899   &9930  &9900  &9910   \\
  $2^3P_0$     & $\chi_{b0}(2P)$        &$10232.5\pm0.4\pm0.5$     &10225.50  &10225  &10233  &10228 &10226 &10235     \\
  $2^1P_1$     & $h_b(2P)$             &$10259.8\pm1.2$            &10248.54  &10250  &10262  &10259 &10247 &10259    \\
  $2^3P_1$     & $\chi_{b1}(2P)$        & $10255.46\pm0.22\pm0.50$ &10244.71  &10246  &10257  &10252 &10244 &10255   \\
  $2^3P_2$     & $\chi_{b2}(2P)$        &$10268.65\pm0.22\pm0.50$  &10258.86  &10261  &10274  &10270 &10257 &10269   \\
  $3^3P_0$     & $\chi_{b0}(3P)$\footnote{not fit}        &                          &10505.30  &10521  &10533  &10496 &10495 &10513  \\
  $3^1P_1$     & $h_b(3P)$             &                           &10523.70  &10540  &10560  &10523 &10591 &10530   \\
  $3^3P_1$     & $\chi_{b1}(3P)$        &$10513.4\pm0.7$           &10520.60  &10537  &10556  &10517 &10580 &10527   \\
  $3^3P_2$     & $\chi_{b2}(3P)$        &$10524.0\pm0.8$           &10532.20  &10549  &10568  &10532 &10578 &10539  \\
  $1^3D_1$     &                        &                          &10142.90  &10137  &10136  &10138 &10112 &10153  \\
  $1^1D_2$     &                        &                          &10152.90  &10148  &       &10146 &10122 &10163  \\
  $1^3D_2$     & $\Upsilon_2(1D)$       & $10163.7\pm1.4$          &10151.95  &10147  &10141  &10145 &10121 &10162  \\
  $1^3D_3$     &                        &                          &10159.40  &10155  &       &10149 &10127 &10170  \\
  $2^3D_1$     &                        &                          &10430.30  &10441  &10454  &10420 &      &10442  \\
  $2^1D_2$     &                        &                          &10438.60  &10450  &       &10428 &      &10450  \\
  $2^3D_2$     &                        &                          &10437.70  &10449  &       &10427 &      &10450  \\
  $2^3D_3$     &                        &                          &10443.90  &10455  &       &10431 &      &10456  \\
  \hline
rel error, $|R|$ (\%)&                  &                          &0.16  & 0.26  & 0.65      & 0.18 & 0.26  & 0.16 \\
  \hline\hline
\end{tabular}
\end{center}
\end{table*}

\begin{table*}[htpb]
\begin{center}
\caption{\label{tab:mass2} Unquenched Bottomonium Mass Spectrum (MeV).}
\footnotesize
\begin{tabular}{ccccccc}
\hline\hline
  $n^{2S+1}L_J$  & States               & RPP~\cite{PDG:2023}      &$M_{0}$ &$\Delta M$  &$M$  \\\hline
  $1^1S_0$     & $\eta_b(1S)$           &$9398.7\pm2.0$            & 9605.28   & -167.17   & 9438.11   \\
  $1^3S_1$     & $\Upsilon(1S)$         &$9460.30\pm0.26$          & 9668.86   & -174.43   & 9494.43   \\
  $2^1S_0$     & $\eta_b(2S)$           &$9999\pm4$                & 10152.45   & -164.90   & 9987.55   \\
  $2^3S_1$     & $\Upsilon(2S)$         &$10023.26\pm0.31$         & 10180.65   & -164.14   & 10016.51   \\
  $3^1S_0$     & $\eta_b(3S)$           &                          &10484.44    &-151.11    &10333.33    \\
  $3^3S_1$     & $\Upsilon(3S)$         &$10355.2\pm0.5$           & 10504.68   & -153.82   & 10350.86   \\
  $4^1S_0$     & $\eta_b(4S)$           &                          &10743.46     &-182.33    &10561.13    \\
  $4^3S_1$     & $\Upsilon(4S)$         &$10579.4\pm1.2$           & 10760.43   & -165.43   & 10595.00    \\
  $5^1S_0$     & $\eta_b(5S)$           &                          &10967.94    &-112.88    &10855.06     \\
  $5^3S_1$     & $\Upsilon(5S)$         &$10885.2^{+2.6}_{-1.6}$   & 10983.08   & -144.15   & 10838.93   \\
  $6^1S_0$     & $\eta_b(6S)$           &                          &11168.80    &-127.23    &11041.57     \\
  $6^3S_1$     & $\Upsilon(6S)$         &$11000\pm4$               & 11183.04   & -123.84   & 11059.20   \\
  $1^3P_0$     & $\chi_{b0}(1P)$        &$9859.44\pm0.42\pm0.31$   & 10034.59  & -176.99   & 9857.60   \\
  $1^1P_1$     & $h_b(1P)$             &$9899.3\pm0.8$            & 10066.90   & -180.2   & 9886.70   \\
  $1^3P_1$     & $\chi_{b1}(1P)$        &$9892.78\pm0.26\pm0.31$   & 10059.46   & -179.86   & 9879.60   \\
  $1^3P_2$     & $\chi_{b2}(1P)$        &$9912.21\pm0.26\pm0.31$   & 10082.15   & -180.46   & 9901.69   \\
  $2^3P_0$     & $\chi_{b0}(2P)$        &$10232.5\pm0.4\pm0.5$     & 10387.65   & -156.18   & 10231.47   \\
  $2^1P_1$     & $h_b(2P)$             &$10259.8\pm1.2$           & 10409.11   & -158.49   & 10250.62   \\
  $2^3P_1$     & $\chi_{b1}(2P)$        & $10255.46\pm0.22\pm0.50$ & 10403.33   & -158.42   & 10244.91   \\
  $2^3P_2$     & $\chi_{b2}(2P)$        &$10268.65\pm0.22\pm0.50$  & 10420.41   & -158.74   & 10261.67   \\
  $3^3P_0$     & $\chi_{b0}(3P)$        &                          &10659.75    &-161.02    &10498.73    \\
  $3^1P_1$     & $h_b(3P)$             &                          &10676.19    &-161.52    &10514.67    \\
  $3^3P_1$     & $\chi_{b1}(3P)$        &$10513.4\pm0.7$           & 10671.38   & -158.86   & 10512.52   \\
  $3^3P_2$     & $\chi_{b2}(3P)$        &$10524.0\pm0.8$           & 10685.47   &-162.89    &10522.58    \\
  $1^3D_1$     &                        &                          &10306.85    &-167.82    &10139.03    \\
  $1^1D_2$     &                        &                          &10316.49    &-167.49    &10149.00    \\
  $1^3D_2$     & $\Upsilon_2(1D)$       & $10163.7\pm1.4$          & 10315.25   & -169.42   & 10145.83   \\
  $1^3D_3$     &                        &                          &10323.07    &-166.98    &10156.09    \\
  $2^3D_1$     &                        &                          &10588.22    &-158.39    &10429.82    \\
  $2^1D_2$     &                        &                          &10596.17    &-158.19    &10437.98    \\
  $2^3D_2$     &                        &                          &10594.94    &-158.42    &10436.52    \\
  $2^3D_3$     &                        &                          &10601.72  &-158.08    &10443.90    \\
  \hline\hline
\end{tabular}
\end{center}
\end{table*}

\begin{table}[htpb]
\caption{Fit Parameters for the Quenched and Unquenched Models.}
\label{gipar}
\begin{center}
\begin{tabular}{ccc}
\hline
\hline
Parameter             &  Quenched  &  Unquenched    \\
\hline
$m_b$                 &  5.1078 GeV        & 4.6878 GeV    \\
$b$                   &  0.1644 GeV$^2$  & 0.1519 GeV$^2$     \\
$c$                   &  -0.4640 GeV      & 0.5431 GeV   \\
$\epsilon_c$          & -0.1600          & -0.2496    \\
$\epsilon_t$          &  0.0278         & 1.3995    \\
$\epsilon_{\mathrm{so(v)}}$  &  -0.0302     & -0.2017    \\
$\epsilon_{\mathrm{so(s)}}$   & 0.0527     & -0.9690    \\
\hline
\hline
\end{tabular}
\end{center}
\end{table}

\begin{table*}[htpb]
\begin{center}
\caption{\label{tab:massshift1} Mass shifts (in MeV) of each channel for the bottomonium.}
\footnotesize
\begin{tabular}{lcccccccccccccccc}
\hline\hline
  Channel  &$1^1S_0$  &$1^3S_1$   &$2^1S_0$   &$2^3S_1$   &$3^1S_0$  &$3^3S_1$   &$4^1S_0$  &$4^3S_1$  &$5^1S_0$  &$5^3S_1$   &$6^1S_0$  &$6^3S_1$  &$1^3P_0$  &$1^1P_1$  &$1^3P_1$  &$1^3P_2$  \\\hline
  $B\bar{B}$            &0.0   &-0.81  &0.0   &-2.66  &0.0    &-4.17  &0.0   &-8.80  &0.0   &-2.46  &0.0   &-2.29  &-4.32  &0.0   &0.0   &-2.47\\
  $B\bar{B^*}$          &-4.33  &-3.27  &-14.96 &-10.40 &-22.60  &-15.71 &-47.02 &-24.01 &-10.27 &-14.58 &-16.34 &-10.85 &0.0   &-10.98 &-9.92  &-7.35\\
  $B^*\bar{B^*}$        &-4.40  &-5.78  &-14.71 &-17.89 &-21.80  &-26.37 &-31.75 &-32.17 &-0.49 &-18.23 &-11.68 &-14.76 &-15.72 &-10.88 &-11.54 &-12.91\\
  $B_s^+B_s^-$          &0.0   &-0.15  &0.0   &-0.33  &0.0    &-0.39  &0.0   &0.34  &0.0   &-0.17  &0.0   &-0.14  &-0.52  &0.0   &0.0   &-0.38\\
  $B_s^+B_s^{*-}$       &-0.84  &-0.61  &-2.0   &-1.36  &-2.31   &-1.56  &-2.16  &-1.33  &-0.59  &-1.52  &-1.37  &-0.88  &0.0   &-1.62  &-1.35  &-1.15\\
  $B_s^{*+}B_s^{*-}$    &-0.87  &-1.10  &-2.02  &-2.41  &-2.34   &-2.75  &-2.14  &-2.33  &-3.12  &-0.89  &-1.52  &-1.71  &-2.57  &-1.65  &-1.89  &-1.83\\
  $B\bar{B}(1^3P_0)$    &-37.34 &0.0   &-27.68 &0.0   &-15.05  &0.0   &-10.51 &0.0   &-8.91  &0.0   &-8.32  &0.0   &0.0   &-35.11 &-0.02  &0.0\\
  $B^*\bar{B}(1^3P_0)$  &0.0   &-38.55 &0.0   &-27.05 &0.0    &-15.03 &0.0   &-10.35 &0.0   &-9.26  &0.0   &-8.11  &-32.73 &-0.02  &-34.02 &-36.05\\
  $B\bar{B}(1^1P_1)$    &0.0   &-13.98 &0.0   &-11.01 &0.0    &-8.34  &0.0   &-7.19  &0.0   &-7.56  &0.0   &-6.72  &-9.32  &0.0   &-16.66 &-12.02\\
  $B^*\bar{B}(1^1P_1)$  &-40.32 &-27.97 &-33.79 &-22.17 &-25.00  &-16.72 &-22.22 &-14.35 &-20.77 &-14.73 &-19.45 &-12.65 &-32.68 &-39.96 &-24.65 &-26.49\\
  $B\bar{B}(1^3P_1)$    &0.0   &-27.11 &0.0   &-19.57 &0.0    &-11.95 &0.0   &-8.92  &0.0   &-8.48  &0.0   &-7.44  &-31.78 &-0.03  &-22.66 &-24.76\\
  $B^*\bar{B}(1^3P_1)$  &-78.15 &-54.12 &-60.81 &-39.53 &-36.73  &-24.22 &-27.71 &-17.97 &-24.09 &-16.81 &-21.75 &-14.49 &-43.38 &-75.01 &-52.54 &-49.68\\
  $B\bar{B}(1^3P_2)$    &-0.18  &-0.13  &-3.01  &-1.68  &-8.16   &-4.29  &-11.08 &-5.33  &-11.96 &-6.58  &-12.37 &-5.80  &0.0   &-1.53  &-1.01  &-0.82\\
  $B^*\bar{B}(1^3P_2)$  &-0.27  &-0.50  &-4.35  &-6.44  &-11.87  &-16.64 &-16.24 &-20.90 &-17.39 &-25.17 &-16.97 &-21.38 &-2.93  &-2.24  &-2.48  &-3.35\\
  $B\bar{B}(2^1S_0)$    &0.0   &-0.04  &0.0   &-0.16  &0.0    &-0.58  &0.0   &-1.09  &0.0   &1.65  &0.0   &-1.54  &-0.50  &0.0   &0.0   &-0.10\\
  $B\bar{B}(2^3S_1)$    &-0.06  &-0.02  &-0.32  &-0.22  &-1.09   &-0.80  &-2.61  &-1.74  &-3.84  &-2.81  &-4.22  &-2.67  &0.0   &-0.25  &-0.43  &-0.05\\
  $B^*\bar{B}(2^1S_0)$  &-0.21  &-0.12  &-0.55  &-0.39  &-1.75   &-1.26  &-3.44  &-2.27  &-4.60  &-3.33  &-4.83  &-3.06  &0.0   &-0.39  &-0.43  &-0.22\\
  $B^*\bar{B}(2^3S_1)$  &-0.20  &-0.17  &-0.71  &-0.87  &-2.41   &-3.07  &-5.46  &-6.36  &-7.83  &-9.93  &-8.42  &-9.35  &-0.54  &-0.54  &-0.27  &-0.85\\
  Total                 &-167.17&-174.43&-164.90&-164.14&-151.11 &-153.82&-182.33&-165.43&-112.88&-144.15&-127.23&-123.84&-176.99&-180.20&-179.86&-180.46\\
  \hline\hline
\end{tabular}
\end{center}
\end{table*}

\begin{table*}[htpb]
\begin{center}
\caption{\label{tab:massshift2} Mass shifts (in MeV) of each channel for the bottomonium.}
\footnotesize
\begin{tabular}{lccccccccccccccccc}
\hline\hline
   Channel  & $2^3P_0$ &$2^1P_1$ &$2^3P_1$ &$2^3P_2$ &$3^3P_0$ &$3^1P_1$ &$3^3P_1$ &$3^3P_2$ &$1^3D_1$ &$1^1D_2$ &$1^3D_2$ &$1^3D_3$ &$2^3D_1$ &$2^1D_2$ &$2^3D_2$ &$2^3D_3$ \\\hline
  $B\bar{B}$            &-10.26 &0.0   &0.0   &-4.34  &-18.88 &0.0   &0.0   &-6.84  &-5.34  &0.0   &0.0   &-4.03  &-9.15  &0.0   &0.0   &-5.43 \\
  $B\bar{B^*}$          &0.0   &-20.67 &-20.35 &-12.66 &0.0   &-29.28 &-28.93 &-17.15 &-5.05  &-17.29 &-16.98 &-10.65 &-8.16  &-24.39 &-24.55 &-13.82\\
  $B^*\bar{B^*}$        &-29.51 &-20.10 &-20.38 &-24.08 &-37.69 &-26.06 &-24.94 &-32.00 &-23.30 &-16.96 &-17.79 &-20.01 &-30.73 &-23.16 &-23.23 &-28.22\\
  $B_s^+B_s^-$          &-0.99  &0.0   &0.0   &-0.49  &-1.03  &0.0   &0.0   &-0.46  &-0.53  &0.0   &0.0   &-0.55  &-0.66  &0.0   &0.0   &-0.54\\
  $B_s^+B_s^{*-}$       &0.0   &-2.32  &-2.16  &-1.49  &0.0   &-2.20  &-2.08  &-1.37  &-0.51  &-2.19  &-2.01  &-1.50  &-0.64  &-2.31  &-2.18  &-1.46\\
  $B_s^{*+}B_s^{*-}$    &-3.64  &-2.34  &-2.52  &-2.68  &-3.43  &-2.21  &-2.30  &-2.55  &-3.34  &-2.22  &-2.46  &-2.39  &-3.38  &-2.33  &-2.47  &-2.61\\
  $B\bar{B}(1^3P_0)$    &0.0   &-19.61 &-0.02  &0.0   &0.0   &-12.08 &-0.01  &0.0   &0.0   &-25.95 &-0.06  &0.0   &0.0   &-14,96 &-0.04  &0.0\\
  $B^*\bar{B}(1^3P_0)$  &-18.48 &-0.02  &-19.23 &-20.47 &-11.06 &-0.01  &-11.64 &-12.85 &-23.74 &-0.06  &-25.56 &-27.87 &-13.51 &-0.04  &-14.61 &-16.30\\
  $B\bar{B}(1^1P_1)$    &-4.13  &0.0   &-12.85 &-8.42  &-6.08  &0.0   &-9.10  &-7.41  &-9.02  &0.0   &-15.24 &-8.78  &-7.49  &0.0   &-10.85 &-7.26\\
  $B^*\bar{B}(1^1P_1)$  &-25.72 &-28.44 &-16.40 &-19.12 &-18.41 &-23.50 &-14.64 &-15.71 &-27.27 &-33.04 &-19.20 &-21.90 &-20.14 &-25.55 &-15.32 &-16.94\\
  $B\bar{B}(1^3P_1)$    &-24.63 &0.03  &-11.23 &-15.33 &-17.43 &-0.02  &-7.16  &-10.65 &-25.43 &-0.09  &-15.42 &-19.05 &-17.30 &-0.06  &-8.57  &-12.40\\
  $B^*\bar{B}(1^3P_1)$  &-21.57 &-45.70 &-34.67 &-29.66 &-13.43 &-30.97 &-23.73 &-20.26 &-33.82 &-57.91 &-43.17 &-37.76 &-20.32 &-36.75 &-28.42 &-23.59\\
  $B\bar{B}(1^3P_2)$    &0.0   &-6.41  &-4.58  &-3.08  &0.0   &-10.41 &-7.59  &-4.79  &-0.79  &-3.89  &-2.79  &-1.93  &-2.04  &-8.91  &-6.53  &-4.06\\
  $B^*\bar{B}(1^3P_2)$  &-14.12 &-9.31  &-10.62 &-13.22 &-24.50 &-15.25 &-17.43 &-21.13 &-7.72  &-5.68  &-6.58  &-8.21  &-18.53 &-13.03 &-15.02 &-18.52\\
  $B\bar{B}(2^1S_0)$    &-0.87  &0.0   &0.0   &-0.48  &-2.43  &0.0   &0.0   &-1.17  &-0.40  &0.0   &0.0   &-0.29  &-1.07  &0.0   &0.0   &-0.96\\
  $B\bar{B}(2^3S_1)$    &0.0   &-0.71  &-0.70  &-0.44  &0.0   &-2.12  &-2.01  &-1.34  &-0.24  &-0.47  &-0.55  &-0.19  &-0.42  &-1.44  &-1.39  &-0.90\\
  $B^*\bar{B}(2^1S_0)$  &0.0   &-1.23  &-1.11  &-0.83  &0.0   &-2.93  &-2.74  &-1.87  &-0.20  &-0.74  &-0.69  &-0.50  &-0.56  &-2.15  &-2.0   &-1.42\\
  $B^*\bar{B}(2^3S_1)$  &-2.27  &-1.61  &-1.62  &-1.98  &-6.65  &-4.48  &-4.58  &-5.34  &-1.15  &-1.02  &-0.94  &-1.38  &-4.29  &-3.13  &-3.24  &-3.67\\
  Total                 &-156.18&-158.49&-158.42&-158.74&-161.02&-161.52&-158.86&-162.89&-167.82&-167.49&-169.42&-166.98&-158.39&-158.19&-158.42&-158.08\\
  \hline\hline
\end{tabular}
\end{center}
\end{table*}

\begin{table*}[htpb]
\begin{center}
\caption{\label{tab:pro1} The probabilities (\%) of each channel for the bottomonium.}
\footnotesize
\begin{tabular}{lcccccccccccccccc}
\hline\hline
  Channel  &$1^1S_0$  &$1^3S_1$   &$2^1S_0$   &$2^3S_1$   &$3^1S_0$  &$3^3S_1$     &$1^3P_0$  &$1^1P_1$  &$1^3P_1$  &$1^3P_2$ & $2^3P_0$ &$2^1P_1$ &$2^3P_1$ &$2^3P_2$ \\\hline
  $B\bar{B}$            &0.00  &0.05  &0.00  &0.33  &0.00  &1.04  &0.47  &0.00  &0.00  &0.21  &2.09  &0.00  &0.00  &0.75  \\
  $B\bar{B^*}$          &0.23  &0.19  &1.65  &1.19  &4.41  &3.35  &0.00  &0.97  &0.95  &0.61  &0.00  &3.41  &3.59  &1.96  \\
  $B^*\bar{B^*}$        &0.23  &0.32  &1.53  &1.92  &3.90  &5.00  &1.19  &0.91  &0.88  &1.17  &4.06  &3.04  &2.85  &3.88  \\
  $B_s^+B_s^-$          &0.00  &0.01  &0.00  &0.03  &0.00  &0.06  &0.04  &0.00  &0.00  &0.03  &0.13  &0.00  &0.00  &0.06  \\
  $B_s^+B_s^{*-}$       &0.04  &0.03  &0.17  &0.12  &0.29  &0.20  &0.00  &0.12  &0.10  &0.08  &0.00  &0.26  &0.26  &0.16  \\
  $B_s^{*+}B_s^{*-}$    &0.04  &0.05  &0.16  &0.20  &0.28  &0.34  &0.17  &0.11  &0.12  &0.14  &0.36  &0.25  &0.25  &0.30  \\
  $B\bar{B}(1^3P_0)$    &1.72  &0.00  &1.78  &0.00  &1.00  &0.00  &0.00  &2.09  &0.00  &0.00  &0.00  &1.28  &0.00  &0.00  \\
  $B^*\bar{B}(1^3P_0)$  &0.00  &1.82  &0.00  &1.68  &0.00  &0.97  &1.82  &0.00  &1.95  &2.12  &1.16  &0.00  &1.22  &1.31  \\
  $B\bar{B}(1^1P_1)$    &0.00  &0.67  &0.00  &0.71  &0.00  &0.62  &0.50  &0.00  &0.99  &0.71  &0.21  &0.00  &0.95  &0.57  \\
  $B^*\bar{B}(1^1P_1)$  &1.79  &1.30  &2.08  &1.37  &1.79  &1.20  &1.83  &2.29  &1.39  &1.54  &1.81  &1.91  &1.06  &1.29  \\
  $B\bar{B}(1^3P_1)$    &0.00  &1.29  &0.00  &1.23  &0.00  &0.81  &1.83  &0.00  &1.30  &1.47  &1.77  &0.00  &0.66  &1.02  \\
  $B^*\bar{B}(1^3P_1)$  &3.46  &2.51  &3.71  &2.41  &2.44  &1.60  &2.35  &4.27  &2.98  &2.87  &1.23  &2.93  &2.30  &1.89  \\
  $B\bar{B}(1^3P_2)$    &0.01  &0.01  &0.20  &0.11  &0.72  &0.39  &0.00  &0.09  &0.06  &0.05  &0.00  &0.53  &0.39  &0.25  \\
  $B^*\bar{B}(1^3P_2)$  &0.01  &0.02  &0.28  &0.43  &0.99  &1.42  &0.14  &0.12  &0.13  &0.19  &1.03  &0.73  &0.79  &1.08  \\
  $B\bar{B}(2^1S_0)$    &0.00  &0.00  &0.00  &0.01  &0.00  &0.05  &0.03  &0.00  &0.00  &0.01  &0.06  &0.00  &0.00  &0.04  \\
  $B\bar{B}(2^3S_1)$    &0.00  &0.00  &0.02  &0.01  &0.09  &0.07  &0.00  &0.01  &0.02  &0.00  &0.00  &0.05  &0.05  &0.04  \\
  $B^*\bar{B}(2^1S_0)$  &0.01  &0.00  &0.03  &0.02  &0.15  &0.11  &0.00  &0.03  &0.02  &0.02  &0.00  &0.10  &0.09  &0.07  \\
  $B^*\bar{B}(2^3S_1)$  &0.01  &0.01  &0.04  &0.05  &0.20  &0.26  &0.04  &0.03  &0.02  &0.05  &0.18  &0.12  &0.13  &0.15  \\
\hline
  Total                 &7.55  &8.28  &11.64 &11.81 &16.35 &17.48 &10.39 &11.05 &10.93 &11.26 &14.10 &14.63 &14.59 &14.81 \\
  $P_{b\bar{b}}$        &92.45 &91.72 &88.34 &88.19 &83.65 &82.52 &89.61 &88.95 &89.07 &88.74 &85.90 &85.37 &85.41 &85.19 \\
  \hline\hline
\end{tabular}
\end{center}
\end{table*}

\begin{table*}[htpb]
\begin{center}
\caption{\label{tab:pro2} The probabilities (\%) of each channel for the bottomonium.}
\footnotesize
\begin{tabular}{lccccccccccccccccc}
\hline\hline
   Channel   &$3^3P_0$ &$3^1P_1$ &$3^3P_1$ &$3^3P_2$ &$1^3D_1$ &$1^1D_2$ &$1^3D_2$ &$1^3D_3$ &$2^3D_1$ &$2^1D_2$ &$2^3D_2$ &$2^3D_3$ \\\hline
  $B\bar{B}$            &13.06  &0.00   &0.00   &3.39  &0.84  &0.00   &0.00   &0.44   &3.31   &0.00   &0.00   &1.31   \\
  $B\bar{B^*}$          &0.00   &11.63  &12.04  &5.67  &0.73  &2.05   &2.19   &1.10   &2.36   &5.89   &6.39   &2.83   \\
  $B^*\bar{B^*}$        &9.22   &7.69   &6.41   &9.74  &2.29  &1.88   &1.87   &2.46   &5.58   &4.84   &4.49   &6.55   \\
  $B_s^+B_s^-$          &0.19   &0.00   &0.00   &0.07  &0.06  &0.00   &0.00   &0.05   &0.12   &0.00   &0.00   &0.07   \\
  $B_s^+B_s^{*-}$       &0.00   &0.34   &0.33   &0.19  &0.06  &0.20   &0.19   &0.13   &0.10   &0.31   &0.31   &0.18   \\
  $B_s^{*+}B_s^{*-}$    &0.43   &0.30   &0.30   &0.36  &0.27  &0.19   &0.20   &0.22   &0.39   &0.29   &0.29   &0.36   \\
  $B\bar{B}(1^3P_0)$    &0.00   &0.67   &0.00   &0.00  &0.00  &1.77   &0.00   &0.00   &0.00   &0.94   &0.00   &0.00   \\
  $B^*\bar{B}(1^3P_0)$  &0.58   &0.00   &0.64   &0.70  &1.55  &0.00   &1.70   &1.87   &0.81   &0.00   &0.90   &1.03   \\
  $B\bar{B}(1^1P_1)$    &0.49   &0.00   &0.68   &0.55  &0.60  &0.00   &1.07   &0.58   &0.58   &0.00   &0.84   &0.52   \\
  $B^*\bar{B}(1^1P_1)$  &1.28   &1.68   &1.07   &1.11  &1.81  &2.18   &1.26   &1.45   &1.48   &1.84   &1.09   &1.22   \\
  $B\bar{B}(1^3P_1)$    &1.24   &0.00   &0.37   &0.67  &1.72  &0.01   &1.01   &1.29   &1.26   &0.00   &0.48   &0.84   \\
  $B^*\bar{B}(1^3P_1)$  &0.64   &1.87   &1.54   &1.19  &2.16  &3.78   &2.85   &2.47   &1.23   &2.38   &1.93   &1.49   \\
  $B\bar{B}(1^3P_2)$    &0.00   &1.02   &0.76   &0.46  &0.06  &0.27   &0.20   &0.13   &0.20   &0.85   &0.64   &0.37   \\
  $B^*\bar{B}(1^3P_2)$  &2.05   &1.41   &1.55   &2.00  &0.45  &0.38   &0.41   &0.59   &1.52   &1.17   &1.30   &1.77   \\
  $B\bar{B}(2^1S_0)$    &0.21   &0.00   &0.00   &0.11  &0.02  &0.00   &0.00   &0.02   &0.09   &0.00   &0.00   &0.09   \\
  $B\bar{B}(2^3S_1)$    &0.00   &0.19   &0.18   &0.12  &0.01  &0.03   &0.04   &0.02   &0.03   &0.12   &0.12   &0.08   \\
  $B^*\bar{B}(2^1S_0)$  &0.00   &0.26   &0.25   &0.17  &0.01  &0.05   &0.05   &0.04   &0.05   &0.19   &0.18   &0.12   \\
  $B^*\bar{B}(2^3S_1)$  &0.55   &0.39   &0.40   &0.47  &0.09  &0.07   &0.07   &0.09   &0.37   &0.27   &0.28   &0.31   \\
\hline
  Total                 &29.94  &27.47  &26.53  &26.97 &12.73 &12.86  &13.13  &12.95  &19.49  &19.12  &19.24  &19.14  \\
  $P_{b\bar{b}}$        &70.06  &72.53  &73.47  &73.03 &87.27 &87.14  &86.87  &87.05  &80.51  &80.88  &80.76  &80.86  \\
  \hline\hline
\end{tabular}
\end{center}
\end{table*}

\begin{figure}
    \centering
    \includegraphics[width=\columnwidth]{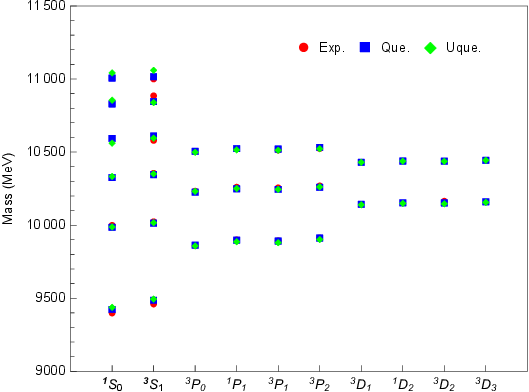}
    \caption{The bottomonium meson spectrum.  ``Exp.'' denotes the current experimental values from the latest PDG~\cite{PDG:2023} and ``Que.'' and ``Uque'' represent the quenched and unquenched model results.}
    \label{fig:spectrum}
\end{figure}

For the $3S$ states, our predicted quenched mass for the  $\Upsilon(3S)$ is about $10346$ MeV, which is close to the experimental data $10355.3\pm 0.5$. There are three well-established vector states reported in the RPP above $B\bar{B}$ threshold, $\Upsilon(4S)$, $\Upsilon(10860)$ and $\Upsilon(11020)$. The last two states are good candidates for the $\Upsilon(5S)$ and $\Upsilon(6S)$~\cite{Wang:2018rjg}, respectively. For the three states,  the predicted masses are 10609 MeV, 10854 MeV and 11016 MeV, respectively in the quenched model, which are close to the experimental data.

The computed unquenched bottomonium spectrum is shown in Table~\ref{tab:mass2}. As we see, the mass shift for each state is generally over 100 MeV. However, the majority of this mass shift can be absorbed by renormalizing the quark model constant term, $c$.
As with the quenched model, the  predicted masses are consistent with experimental results.

Many years ago, Geiger and Isgur noted that mass shifts due to individual continuum channels can appear to be quite random, however the sum over spin multiplets tend to look similar~\cite{Geiger:1991ab}. This observation was subsequently proven to be true for a large class of pair creation models in the limit of degenerate channel multiplets~\cite{Barnes:2007xu}. It is for this reason that the computation presented here sums over combinations of the 1S, 1P, and 2S multiplets.  Tables~\ref{tab:massshift1} and \ref{tab:massshift2} report mass shifts for 32 bottomonium states for each of the 18 channels considered here and reveal details concerning this observation. One sees, for example, that total mass shifts tend to be approximately 160 MeV, in keeping with the notion of the the ``effective renormalisability" of the unquenched model. Furthermore, the S+P-wave channel tends to cause large mass shifts with respect to the S+S-wave channels, with the dominance decreasing as one moves higher in the spectrum. Lastly, the 1S+2S channel typically contributes approximately 10\% of the shift due to the 1S+1S channel. This is an indication that the multiplet expansion considered here converges rapidly for the states considered, thereby lending support to the methodology employed.

\subsection{Continuum Probabilities and Upsilon Decay Constants}

Tables \ref{tab:pro1} and \ref{tab:pro2}  report the probabilities associated with each continuum channel in accord with Eq. \ref{deltam2}. These probabilities tend to be at the percent or permil level, and there is evidence of convergence as one moves up channel quantum numbers. The resulting bare quark-antiquark components comprise between 70\% and 92\% of the physical states for the 26 mesons considered.  This observation raises interesting questions about the viability of the naive constituent quark model; for example, the strength of the meson decay constant is driven by the wavefunction at the origin (cf. the classic work by van Royen and Weisskopf~\cite{VanRoyen:1967nq}), implying that the unquenched quark model predictions for this quantity should be lessened by 70-90\%.

We have tested this effect using a relativistic extension of the method of van Royen and Weisskopf reported in Ref.~\cite{Lakhina:2006vg}. The results are presented in Table~\ref{tab:decayConstant} for a nonrelativistic quark model (Ref.~\cite{Barnes:2005pb}) and the Godfrey-Isgur model used here. We note the over-estimate of decay constants that is typical of nonrelativistic quark models in column two. The momentum factors of the Godfrey-Isgur model tend to soften the wavefunction at the origin, leading to the smaller results of column four.  In both cases, the wavefunction renormalization that happens due to unquenching has the welcome effect of bringing the predictions closer to experiment.

We remark that this is not the end of the story: the unquenched quark model is a (nonrelativistic) field theory, and in principle \textit{all} the parameters appearing in the model should be renormalized.  For example, setting $\alpha = 1.2 \alpha_0$, where $\alpha_0$ is the bare fine structure constant, brings the nonrelativistic decay constants into very good agreement with experiment. A final warning concerns the experimental values of the decay constants for higher excitations. A recent coupled channel K-matrix analysis of  bottomonium production indicates that extracted $e^+e^-$ couplings are sensitive to modeling details and that commonly employed experimental techniques such as using the Breit-Wigner amplitude model and ratios of cross sections to determine partial widths can lead to incorrect results~\cite{Husken:2022yik}. Thus overly precise comparison of predictions to PDG values for decay constants can be misleading.

\begin{table*}[ht]
\begin{tabular}{c|cc|cc|c}
\hline\hline
state & \multicolumn{2}{c|}{$f_\Upsilon$ NonRel (MeV)}  & \multicolumn{2}{|c|}{$f_\Upsilon$ GI (MeV)} & RPP \\
     & quenched & unquenched  & quenched & unquenched &  \\
\hline
$\Upsilon$  & 963 & 886 & 740 & 680 & 708(8) \\
$\Upsilon'$  & 640 & 563 & 510 & 450 & 482(10) \\
$\Upsilon''$  & 555 & 460 & 435 & 360  & 346(50) \\
\hline\hline
\end{tabular}
\caption{Quenched and Unquenched Vector Decay Constants.}
\label{tab:decayConstant}
\end{table*}

\subsection{Perturbative Mass Shifts and the Robustness of the Quark Model}

Constructing unquenched quark models such as that presented here can shed light on the rather perplexing issue of how quenched models can possibly represent hadronic physics with any accuracy when coupled channel effects are so strong. (This problem goes back at least 60 years, with Oakes and Yang wondering why coupled channel effects do not ruin the Gell-Mann--Okubo mass formula~\cite{Oakes:1963zz}).
In other words, why is the (constituent) quark model so robust? As we have mentioned, there is a theorem which states that mass shifts become equal when sums over multiplet members are made in the degenerate multiplet mass limit. This certainly makes the quark model robustness plausible since it is trivial to absorb constant mass shifts in this case.

Nevertheless, channel multiplets are not degenerate and it is of interest to examine whether the quark model remains robust more generally. Of course, the numerical results presented here imply that it does indeed remain so (at least for the particular model implementations chosen). This issue can be examined in more detail by considering the case of small perturbations from the degenerate multiple mass limit. We consider the case of coupling to a nS-mS channel and assume that the  multiplet member masses are perturbed according to the (bare) quark model. Thus one has

\begin{eqnarray}
\delta M_0 &=& -\frac{3}{4} \Delta_h, \nonumber \\
\delta M_1 &=& +\frac{1}{4} \Delta_h,
\end{eqnarray}
where $\Delta_h$ is the expectation value of the spatial part of the hyperfine interaction in the unperturbed nS multiplet and the subscripts refer to the singlet and triplet members of the multiplet.

The parametric dependence of the perturbative mass on the multiplet masses arises from the reduced mass, $\mu_{BC}$ and the energy denominator, $M_0 - M_B - M_C$, where we have denoted the members of a nS-mS channel as $B$ and $C$. Both of these depend, in turn, on the sum $\delta_B + \delta_C$ (where $\delta_B$ is the mass shift for meson $B$). Shifts in the multiplet wavefunctions are higher order. We thus conclude that the perturbative correction to the mass shift for each meson coupling to a given multiplet is proportional to

\begin{equation}
\Delta M_A = \sum_{BC \in \text{multiplet}} (\delta_B + \delta_C)\, \Delta M_{BC},
\label{eqn:shift}
\end{equation}
with the latter mass shift defined in Eq.~\ref{dm}. For the S-wave multiplet channels considered here $\delta_B+\delta_C$ is given by $-3\Delta_h/2$, $-\Delta_h/2$, and $\Delta_h/2$ for the channels $(BC) = (00)$,
$(01)$, and $(11)$ respectively. Alternatively,
the factorisable form of the decay model implies that, for a given multiplet,  the  matrix elements appearing in Eq. \ref{eqn:shift} are related by Clebsch-Gordon coefficients. For example, these coefficients are $(1,4,7)$ for $n^3S_1$ decay and $(0,6,6)$ for $n^1S_0$ decay. Substituting these ratios into Eq.~\ref{eqn:shift} then reveals the
 surprising result that there are no additional mass shifts for nS bottomonium multiplets when coupled to a perturbatively shifted S-wave multiplet.

For P-wave mesons the relevant spin coefficients for S-wave decay are

\begin{eqnarray}
^3P_2 &=& (0,0,2),\ \ ^3P_1 = (0,2,0), \nonumber \\
^3P_0 &=& (3/2,0,1/2), \ \ ^1P_1 = (0,1,1),
\end{eqnarray}
again in the order (00), (01), (11). Substituting into Eqn.~\ref{eqn:shift} gives perturbative mass shifts for the mesons of
\begin{equation}
\Delta[M(^3P_2), M(^3P_1), M(^3P_0), M(^1P_1)]_S = [1,-1,-2,0] \Delta_h.
\end{equation}
Remarkably, these are precisely the mass splittings induced by a spin-orbit interaction. We conclude that unquenching P-wave mesons with nS+mS coupled  channels induces an effective spin-orbit interaction in the quenched quark model, which can therefore be completely renormalised. This story repeats for D-wave decay amplitudes in the P-wave multiplet. In this case the spin coefficients are

\begin{eqnarray}
^3P_2 &=& (1,3,8/3)\ \ ^3P_1 = (0,5/3,5), \nonumber \\
^3P_0 &=& (0,0,20/3) \ \ ^1P_1 = (0,10/3,10/3),
\end{eqnarray}
and the mass shifts are
\begin{equation}
\Delta[M(^3P_2), M(^3P_1), M(^3P_0), M(^1P_1)]_D = -\frac{5}{3}[1,-1,-2,0] \Delta_h.
\end{equation}
Thus spin-orbit-like perturbations also occur in the D-wave mass splitting amplitudes.

Because the channel mass splittings are generated by spin-dependent hyperfine, spin-orbit, and tensor interactions, and the decay channels are also related by spin Clebsch-Gordon coefficients, one expects the effective interaction created by unquenching should be describable in terms of general spin-dependent perturbations. Of course some of these may not be present in the spin-dependent interactions generated by one gluon exchange (which typically inform quark models) and the  spatial dependence of the operators change; nevertheless, the results presented here make it likely that much of unquenching effects can indeed be absorbed by quark model parameters, thereby motivating the robustness of the quark model under unquenching and underpinning the utility of the constituent quark approach  to hadronic structure.

\section{Summary}
\label{sec:summary}

The Godfrey-Isgur relativised quark model has been used to compute a range of bottomonium masses, yielding excellent agreement with the measured spectrum. The model has been ``unquenched" by incorporating coupling to 1S-1S, 1S-1P, and 1S-2S continuum channels via the $^3P_0$ hadronic decay model vertex. These calculations have been performed with exact unperturbed wavefunctions, which is an improvement over previous work that relied on approximate simple harmonic oscillator wavefunctions. The resulting spectrum is also in excellent agreement with experiment.

These results motivate a resolution to the Oakes-Yang problem: namely the constituent quark model is capable of absorbing the effects of unquenching into the model itself. Or, restating, the model space spanned by typical spin-dependent quark models is sufficient to largely  absorb the effects of unquenching.

These numerically determined conclusions were verified in the case of perturbative mass shifts of S- and P-wave states coupled to nS-mS channels and it was argued that this should be a general feature. Of course the spatial part of the effective operators will also shift and new exotic spin-dependence may emerge, thus the renormalisability of the unquenched model will only be approximate. Indeed, shifts in hadronic properties can be significant near threshold openings, where our arguments do not hold.

Finally, we have examined the effect of unquenching on vector bottomonium decay constants, where the flow of probability into the continuum is expected to play an important role. We find that the low lying decay constants drop in magnitude, and that they do so by an increasing amount higher in the spectrum. Both effects are welcome, in that they bring the predictions of the unquenched quark model closer to experiment.

The importance of unquenching has been appreciated in hadronic physics for many decades, and much work has been done investigating its effects. Much remains to be done. For example, a full renormalization procedure has yet to be carried out. Also, the sum over intermediate states is only roughly addressed in computations to date, and must be investigated, perhaps with new methods. Joining the approach to QCD in the high energy regime and incorporating multi-pion intermediate states also remain open problems.

\section{Acknowledgements}
Swanson acknowledges support by the U.S. Department of Energy under contract DE-SC0019232. This work contributes to the goals of the US DOE ExoHad Topical Collaboration, Contract No. DE-SC0023598.

\bibliographystyle{unsrt}
\bibliography{cite}  

\end{document}